\documentclass{PoS}

\usepackage{graphicx}
\usepackage{amssymb}
\usepackage{siunitx}
\usepackage{pdfpages}
\usepackage{subfigure}
\usepackage{float}
\usepackage{amsmath}
\usepackage{color}
\usepackage{ulem}
\usepackage{lineno}

\usepackage{xspace}
\usepackage{booktabs}
\usepackage{multirow}

\newcommand{\figref}[1]{Fig.~\ref{#1}}
\newcommand{\Figref}[1]{Figure~\ref{#1}}
\newcommand{\secref}[1]{Sec.~\ref{#1}}

\newcommand{\NeCOtwoNtwo}{Ne-CO$_2$-N$_2$ (90-10-5)\xspace}

\newcommand{\ibf}{ion backflow\xspace}

\title{Discharge probability studies with multi-GEM detectors for the ALICE TPC Upgrade}

\ShortTitle{Discharge probability of multi-GEM structures.}

\author{\speaker{P. Gasik} for the the ALICE TPC collaboration\\
        Physik Department E62, Technische Universit\"at M\"unchen, Garching.\\
        Excellence Cluster 'Origin and Structure of the Universe', Garching.\\
        E-mail: \email{p.gasik@tum.de}}

\abstract{
A large Time Projection Chamber (TPC) is the main device for tracking and charged-particle identification in the ALICE experiment at the CERN LHC. After the second long shutdown in 2019-2020, the LHC will deliver Pb beams colliding at an interaction rate of up to 50 kHz, which is about a factor of 50 above the present readout rate of the TPC. To fully exploit the LHC potential, the TPC readout chambers will be upgraded with Gas Electron Multiplier (GEM) technology. 

To assure stable behaviour of the upgraded chambers in the harsh LHC environment, a dedicated R\&D programme was launched in order to optimize GEM stack geometry and its high voltage configuration with respect to electric discharges. We present a summary of discharge probability measurements performed with 3- and 4-GEM prototypes irradiated with highly ionising alpha particles.
}

\FullConference{5th International Conference on Micro-Pattern Gas Detectors (MPGD2017)\\
		22-26 May, 2017\\
		Philadelphia, USA}

\begin{document}
\section{Introduction}
The ALICE Collaboration is planning a major upgrade of the detector apparatus during the second LHC long shutdown (LS2, 2019-20) in order to fully exploit the LHC potential in Runs 3 and 4 (2021-2029) \cite{LOI}. The cornerstone of the project is the upgrade of the ALICE Time Projection Chamber (TPC) \cite{TPCU}. The currently operated Multi Wire Proportional Chambers will be replaced by the detectors based on the Gas Electron Multiplier (GEM) \cite{Sauli} technology.

In the intensive  R\&D  programme, summarised in \cite{TDR,Addendum}, a quadruple GEM (4-GEM) amplification scheme has been identified as a baseline solution for the upgrade. The new chambers will employ stacks containing Standard (S, 140\,$\mu$m pitch) and Large Pitch (LP, 280\,$\mu$m pitch) GEMs in the configuration S-LP-LP-S. This solution fulfils challenging requirements of the upgrade in terms of energy resolution, ion backflow (IBF) and stability against electrical discharges \cite{Addendum,Dedx}.

The latter is in particular important as the application of GEM technology in the ALICE TPC entail unprecedented challenges in terms of expected load in 50\,kHz Pb-Pb collisions. Up to now, however, the only comprehensive, phenomenological discharge study with single, double and triple GEM detectors has been reported in \cite{Bachmann} and concerns mainly Ar-based gas mixtures. Therefore, a dedicated R\&D was launched in order to establish a safe working point of the GEM chambers in terms of their geometrical and HV configuration. This manuscript summarizes 3- and 4-GEM studies, following the discussions in \cite{Addendum}.
\section{Experimental setup}
\label{sec:setup}
\begin{figure}[ht]                                                                                                             
  \centering                                                                                                                   
  \includegraphics[width=0.48\textwidth]{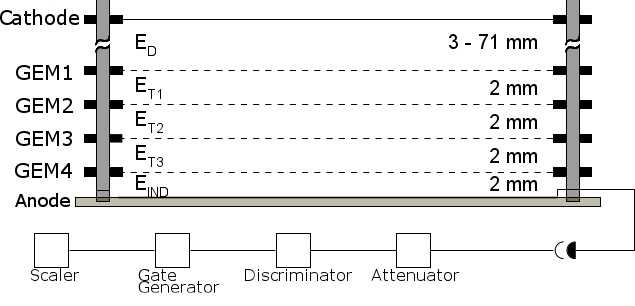}                                                          
  \caption[Experimental setup]{Scheme of the experimental setup (4-GEM configuration).}        
  \label{fig:pgasik:setup}                                                                                                   
\end{figure}

The scheme of the experimental setup is shown in \figref{fig:pgasik:setup}.
The detector housing contains a 10$\times$10\,cm$^2$ GEM holder, a drift cathode and a readout anode. Single- and multi-GEM stacks can be installed. The setup does not employ a field cage giving flexibility to adjust the drift gap length (distance between the cathode and the GEM stack) continuously between 3 and 71\,mm in case of a 4-GEM configuration. 

High voltage is applied to the GEM stack via a resistor chain which defines the potential on each GEM electrode. The gain of the setup at given HV settings is determined by the usual method of recording the current at the pad plane and the rate of absorbed X-rays of known energy (e.g. 5.9\,keV X-rays from an $^{55}$Fe source).

The occurrence of a spark in a GEM foil is detected according to the readout scheme presented in \figref{fig:pgasik:setup}. A raw signal induced on the pad plane is attenuated (1-31 dB) and directed into a discriminator unit which filters out low-amplitude signals ($\sim$10\,mV) induced by alpha particles and triggers on large-amplitude ($>$1\,V) discharge signals. Since the raw signals are often modified by the RLC response of the system (signal oscillations), a gate is created when the discriminator threshold is exceeded which is then counted by a scaler. This way, multi-counting of the same signal can be avoided. 

The discharge probability is defined as the ratio of the number of detected discharges to the total number of particles irradiating the detector. For the studies presented in this report, the detector was irradiated with highly ionising alpha particles emitted with a rate of $\sim0.5$\,Hz from an internal, gaseous $^{220}$Rn source or with a rate of $\sim600$\,Hz from a collimated mixed $^{239}$Pu+$^{241}$Am+$^{244}$Cm source shooting through an 8\,mm hole in the cathode, perpendicular to the GEM stack. The energy of the alpha particles varies between 5.2 and  6.4\,MeV, depending on the nuclide.

The discharge studies presented in this work have been performed in Ar-CO$_2$ (90-10) and Ne-CO$_2$-N$_2$ (90-10-5), the latter being the ALICE TPC mixture.

Two groups of HV settings were used for the measurements, the so-called "standard" and "IBF" settings. The former are typically used with conventional 3-GEM systems operated in high-rate environment and are optimized in terms of stability by decreasing the voltage across GEMs (and thus their gain) towards the bottom of the stack \cite{Bachmann}. The second group of settings is optimized for low ion backflow and allowing their usage in TPCs (e.g ALICE \cite{TDR, Addendum}), thus the gain of the bottom foils is higher than those on the top of a stack. This, however, may compromise the overall stability of the system. Each setting can be scaled in order to vary the total gain. The nominal drift field of the ALICE TPC, $E_{\mathrm{D}}=400\,\mathrm{V}/\mathrm{cm}$, is applied in all measurements.
\section{Studies with a 3-GEM setup}
\label{sec:3gem}
\begin{figure}[]
\centering
   \includegraphics[width=0.49\linewidth]{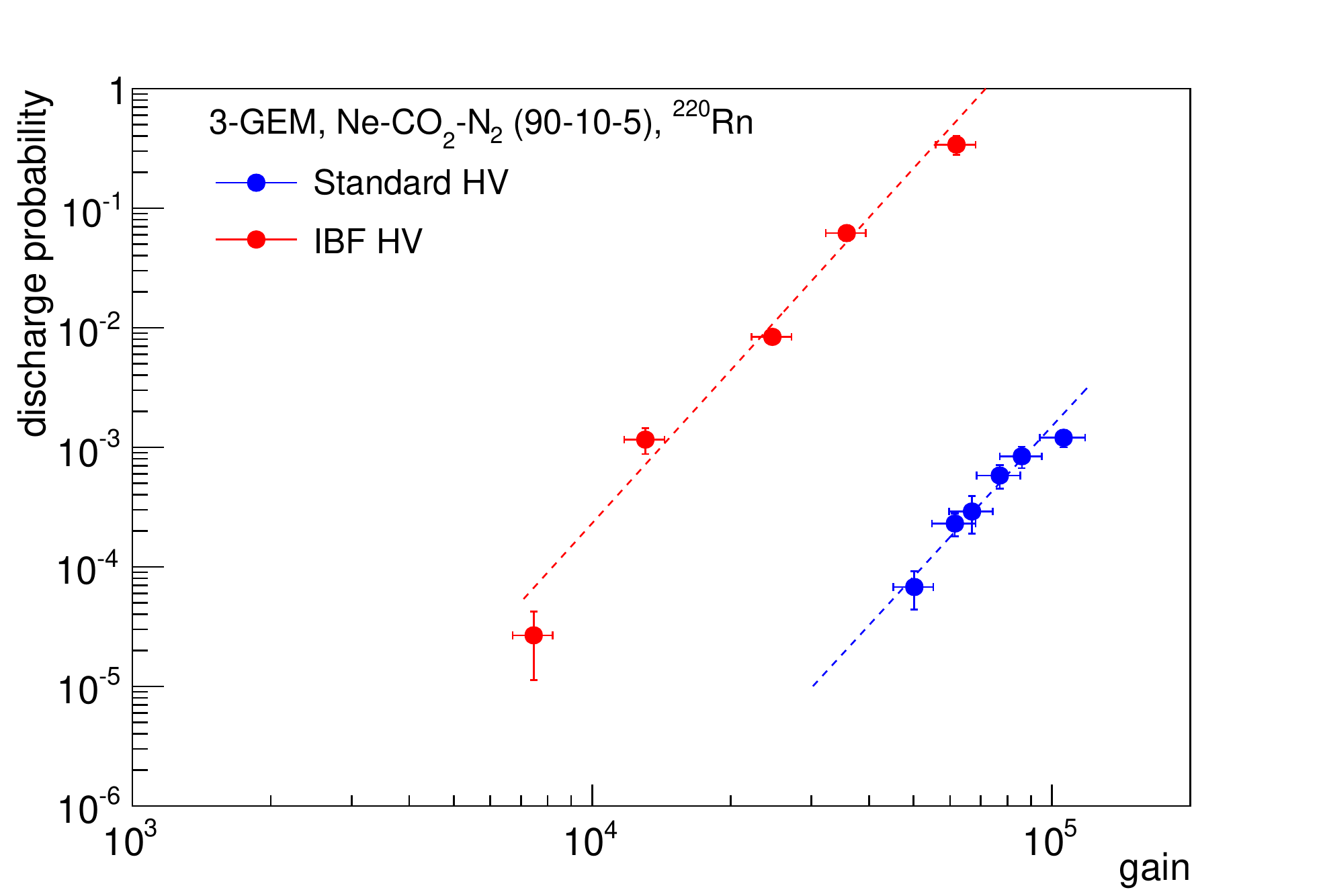}
   \includegraphics[width=0.49\linewidth]{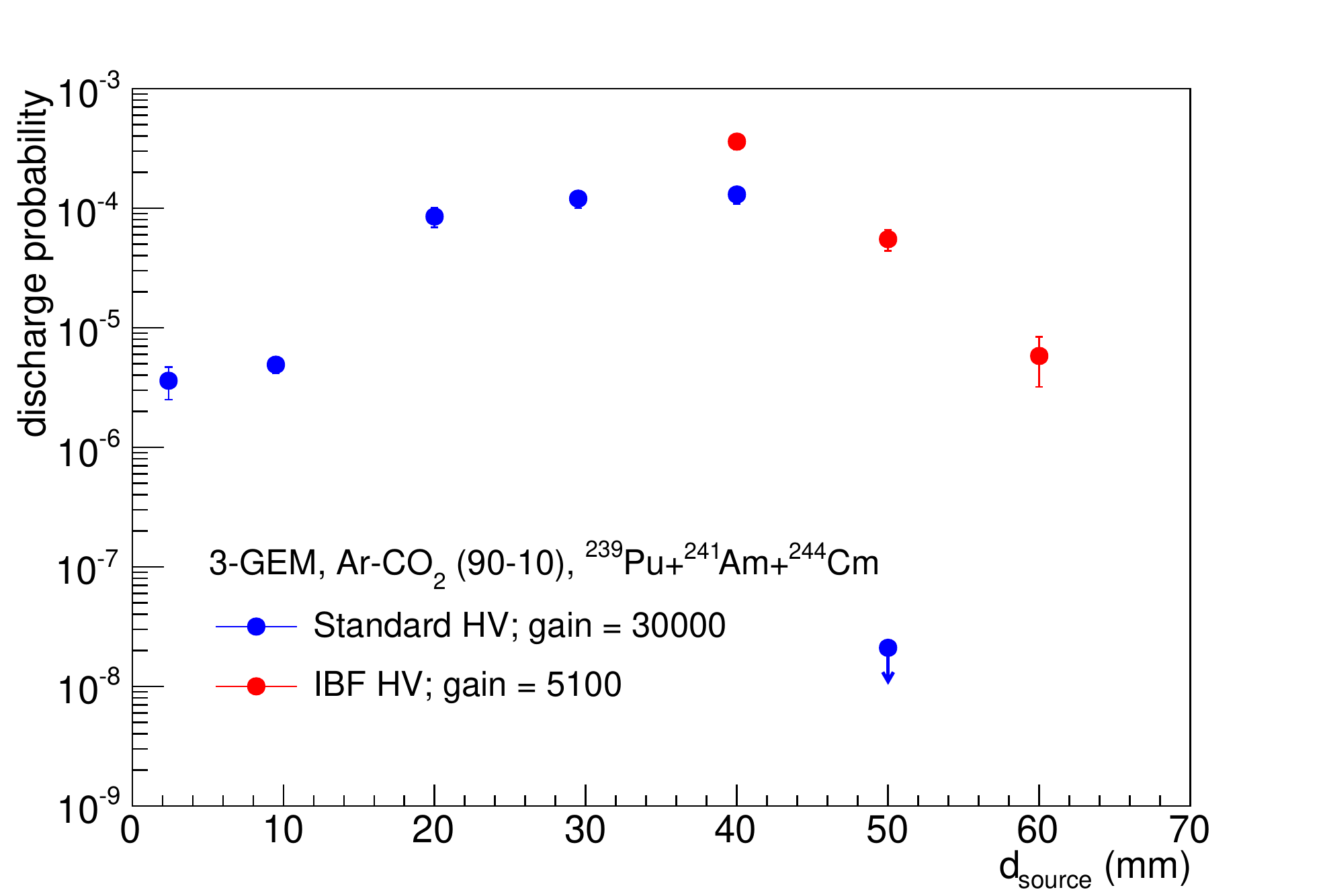}
  \caption[Discharge probability of a triple GEM detector]{Left: discharge probability of a triple GEM detector measured for different HV settings (see text). Dashed lines represent power law function fits. Right: discharge probability of a triple GEM detector measured as a function of the distance between the source and the GEM stack with the mixed nuclide source for different HV settings. Upper limits for discharge probability are indicated with arrows.}
  \label{fig:3gemhv}
\end{figure}

The discharge probability in a 3-GEM setup is measured as a function of the effective gain for different HV settings applied to the stack. 
\Figref{fig:3gemhv} (left panel) shows the results obtained in \NeCOtwoNtwo  using the $^{220}$Rn source. Due to the low rate of the source, the measurements were performed at gains higher than the envisaged nominal ALICE TPC gas gain of 2000.
 In addition to the standard settings, used as a reference, the detector was also operated in the low IBF configuration. 
The latter yields a noticeable decrease in detector
stability of more than three orders of magnitude, which underlines the strong dependence of the detector stability on the HV settings. In order to get an estimate of the discharge probability at the nominal gas gain,
the data points are fitted with power law functions and extrapolated to the  gain of 2000,
which yields  ${\sim}10^{-10}$ and ${\sim}2\times10^{-7}$ for the standard and the IBF
settings, respectively. This clearly indicates that the stability of the IBF-optimized 3-GEM stack is not sufficient.

The deterioration of the detector stability can be also observed in studies with the collimated alpha source shooting perpendicular to the GEMs through the hole in the cathode (see \secref{sec:setup}). Right panel of \figref{fig:3gemhv} shows the discharge probability as a function of the distance $d_{\mathrm{source}}$ between the source and the GEM stack measured in the Ar-CO$_2$ (90-10) gas mixture. The broad plateau in measurements with the standard settings clearly indicates that the discharge probability is higher when the charge deposit occurs in the closest vicinity of the GEM holes. When the distance is increased such that the alphas do not longer reach the GEM structure (assuming range of alpha particles in the mixture of $\sim 4.8$\,cm \cite{Density}) the discharge probability drops by several orders of magnitude. This is due to the fact that the density of the charge that arrives at the GEM holes is reduced by diffusion and will therefore have less probability to trigger a spark.
This simple geometrical explanation has been recently confirmed with single-GEM measurements followed  by GEANT4 simulations \cite{Density}.

\Figref{fig:3gemhv} shows also the influence of changing the HV settings applied to the stack from standard to IBF-optimised. First of all, the discharge probability measured for $d_{\mathrm{source}}=40$\,mm is similar in both measurements, even though the detector with the IBF settings was operated at a much lower gain than in the standard configuration. Secondly, the sudden drop of discharge probability for distances larger than the alpha range is diluted in case of the IBF settings. A measurable discharge probability is obtained even for distances exceeding the range of alphas. This indicates that the reduced stability is related to the higher discharge rate in the bottommost GEM of the stack, where the amplification is the highest.

The track length scan indicates that the development of a spark in a GEM foil is influenced mainly by the local charge deposited in a single hole rather than total charge integrated over the whole foil area. We conclude that it is the number of particles that cross the GEM stacks, liberating charge in the closest vicinity of the GEM holes, which determine the discharge rate of the detectors. The primary charge that reaches the readout chambers from the drift volume has significantly less impact on the detector stability.
\section{Studies with a 4-GEM setup}
The stability tests of a 4-GEM detector have been performed for different types of foils in the stack, following the \ibf and energy resolution studies performed in the scope of the ALICE TPC upgrade R\&D \cite{TDR, Addendum}. The gas mixture used in these measurements is always \NeCOtwoNtwo. In measurements with the high-rate alpha source the drift gap is set to 38\,mm to assure all alpha particles cross the GEM plane which can be considered as the worst-case scenario for the detector stability (see \secref{sec:3gem}).

\begin{figure}[]
  \centering
   \includegraphics[width=0.49\linewidth]{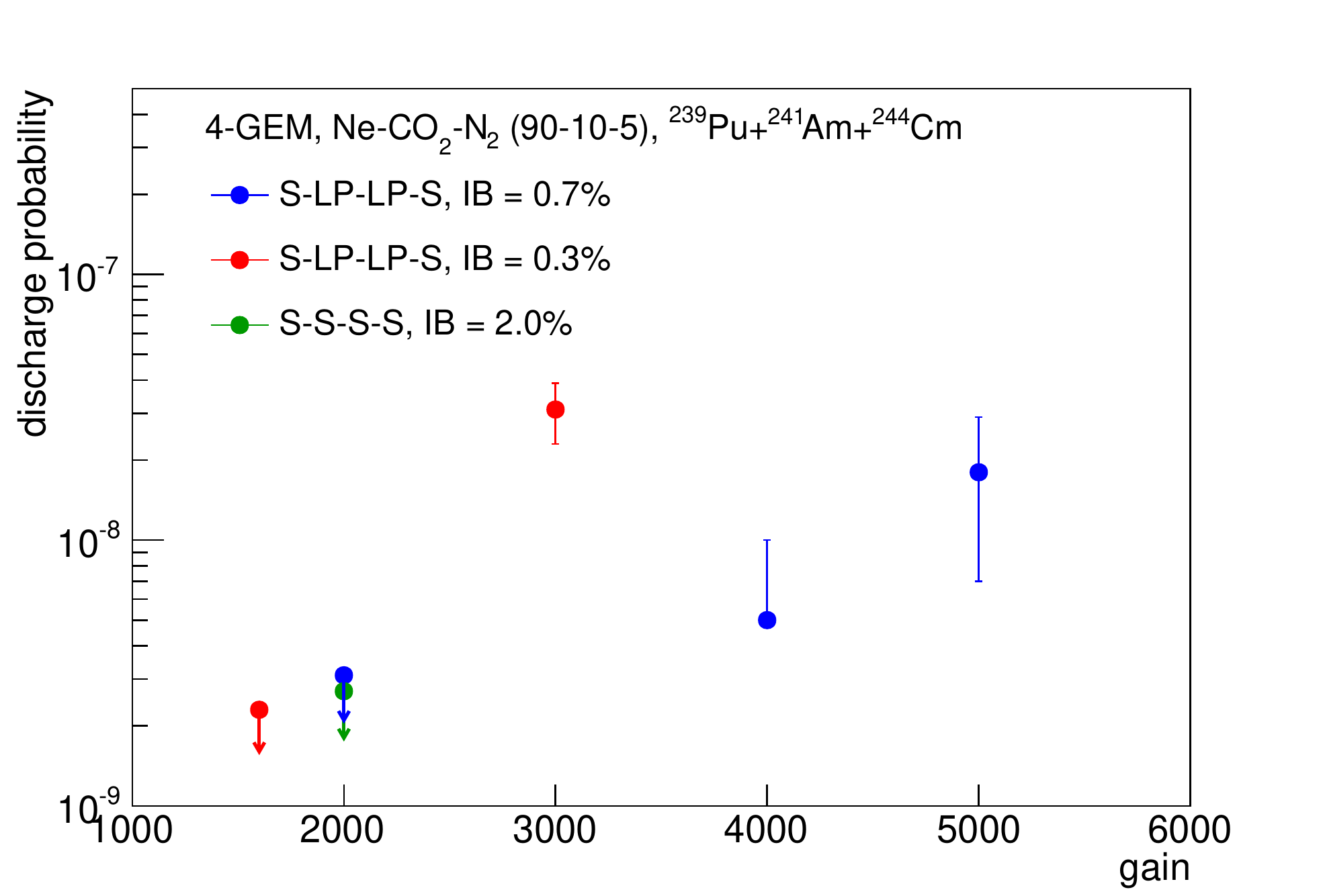}
  \caption[Discharge probability of quadruple GEM detectors]{Discharge probability of a quadruple GEM prototype measured for different stack and HV configurations. Corresponding values of \ibf are indicated. Upper limits for discharge probability are indicated with arrows.}
  \label{fig:4gemgain}
\end{figure}

\Figref{fig:4gemgain} presents the discharge probability measured for various 4-GEM stack configurations. At a gain of $\sim$2000 only upper limits of the probability are indicated which means that during the time of measurement at a given setting no discharge was recorded. The results show that any of the 4-GEM configurations is more stable than the standard triple GEM operated in the low ion backflow mode (see \secref{sec:3gem}). 

The S-LP-LP-S configuration has been checked with two different HV configurations: the so-called ALICE "baseline", optimised for both \ibf ($IB\approx0.7\%$) and energy resolution of $^{55}$Fe X-ray peak ($\sigma_{\mathrm{55Fe}}\approx12\%$), and one with a very low \ibf value ($IB\approx0.3\%$) but worse energy resolution of $\sigma_{\mathrm{55Fe}}\approx17\%$. The comparison of both settings at the gain of $\sim$3000 shows that tuning the settings to obtain the lowest possible value of the IBF seriously affects the stability of the system.
\section{Summary}
The discharge probability of triple and quadruple GEM stacks has been measured in the course of the ALICE TPC upgrade R\&D programme. In the 3-GEM stack, at a gas gain of 2000 in the \NeCOtwoNtwo mixture, a discharge probability of about $10^{-10}$ is estimated in the standard HV configuration. This can be considered as safe for GEM operation in a high-rate environment. Optimization of the voltage settings with respect to minimal \ibf, however, leads to an increase of the discharge probability by more than three orders of magnitude. 

To fulfil the challenging requirements of the upgrade, the readout chambers will be based on the 4-GEM stack configuration. In this work we show that the additional GEM layer leads to a more stable operation even with the IBF-optimised HV settings. An upper limit of the discharge probability of 3.1$\times$10$^{-9}$ per alpha particle has been achieved. The stability studies indicate (in addition to the \ibf and energy resolution optimisation) that the S-LP-LP-S configuration of the readout chambers is suitable for the TPC upgrade purposes.
\newpage
\section*{Acknowledgements}
The author acknowledges support by the DFG Cluster of Excellence "Origin and Structure
of the Universe" (www.universecluster.de) [project number DFG EXC153] and by the Federal Ministry of Education and Research (BMBF, Germany) [grant number 05P15WOCA1].

\end{document}